\documentstyle[twoside,graphicx]{article} 

\catcode`\@=11
\long\def\@makefntext#1{
\protect\noindent \hbox to 3.2pt {\hskip-.9pt  
$^{{\eightrm\@thefnmark}}$\hfil}#1\hfill}		

\def\@makefnmark{\hbox to 0pt{$^{\@thefnmark}$\hss}}	
	
\def\ps@myheadings{\let\@mkboth\@gobbletwo
\def\@oddhead{\hbox{}
\rightmark\hfil\eightrm\thepage}   
\def\@oddfoot{}\def\@evenhead{\eightrm\thepage\hfil
\leftmark\hbox{}}\def\@evenfoot{}
\def\sectionmark##1{}\def\subsectionmark##1{}}

\oddsidemargin=\evensidemargin
\newcounter{sectionc}\newcounter{subsectionc}\newcounter{subsubsectionc}
\renewcommand{\section}[1] {\vspace{12pt}\addtocounter{sectionc}{1} 
\setcounter{subsectionc}{0}\setcounter{subsubsectionc}{0}\noindent 
	{\tenbf\thesectionc. #1}\par\vspace{5pt}}
\renewcommand{\subsection}[1] {\vspace{12pt}\addtocounter{subsectionc}{1} 
	\setcounter{subsubsectionc}{0}\noindent 
	{\bf\thesectionc.\thesubsectionc. {\kern1pt \bfit #1}}\par\vspace{5pt}}
\renewcommand{\subsubsection}[1] {\vspace{12pt}\addtocounter{subsubsectionc}{1}
	\noindent{\tenrm\thesectionc.\thesubsectionc.\thesubsubsectionc.
	{\kern1pt \tenit #1}}\par\vspace{5pt}}
\newcommand{\nonumsection}[1] {\vspace{12pt}\noindent{\tenbf #1}
	\par\vspace{5pt}}

\newcounter{appendixc}
\newcounter{subappendixc}[appendixc]
\newcounter{subsubappendixc}[subappendixc]
\renewcommand{\thesubappendixc}{\Alph{appendixc}.\arabic{subappendixc}}
\renewcommand{\thesubsubappendixc}
	{\Alph{appendixc}.\arabic{subappendixc}.\arabic{subsubappendixc}}

\renewcommand{\appendix}[1] {\vspace{12pt}
        \refstepcounter{appendixc}
        \setcounter{figure}{0}
        \setcounter{table}{0}
        \setcounter{lemma}{0}
        \setcounter{theorem}{0}
        \setcounter{corollary}{0}
        \setcounter{definition}{0}
        \setcounter{equation}{0}
        \renewcommand{\thefigure}{\Alph{appendixc}.\arabic{figure}}
        \renewcommand{\thetable}{\Alph{appendixc}.\arabic{table}}
        \renewcommand{\theappendixc}{\Alph{appendixc}}
        \renewcommand{\thelemma}{\Alph{appendixc}.\arabic{lemma}}
        \renewcommand{\thetheorem}{\Alph{appendixc}.\arabic{theorem}}
        \renewcommand{\thedefinition}{\Alph{appendixc}.\arabic{definition}}
        \renewcommand{\thecorollary}{\Alph{appendixc}.\arabic{corollary}}
        \renewcommand{\theequation}{\Alph{appendixc}.\arabic{equation}}
        \noindent{\tenbf Appendix \theappendixc #1}\par\vspace{5pt}}
\newcommand{\subappendix}[1] {\vspace{12pt}
        \refstepcounter{subappendixc}
        \noindent{\bf Appendix \thesubappendixc. {\kern1pt \bfit #1}}
	\par\vspace{5pt}}
\newcommand{\subsubappendix}[1] {\vspace{12pt}
        \refstepcounter{subsubappendixc}
        \noindent{\rm Appendix \thesubsubappendixc. {\kern1pt \tenit #1}}
	\par\vspace{5pt}}

\newcommand{\textlineskip}{\baselineskip=13pt}
\newcommand{\smalllineskip}{\baselineskip=10pt}

\def\eightcirc{
\begin{picture}(0,0)
\put(4.4,1.8){\circle{6.5}}
\end{picture}}
\def\eightcopyright{\eightcirc\kern2.7pt\hbox{\eightrm c}}

\def\abstracts#1#2#3{{
	\centering{\begin{minipage}{4.5in}\baselineskip=10pt\footnotesize
	\parindent=0pt #1\par 
	\parindent=15pt #2\par
	\parindent=15pt #3
	\end{minipage}}\par}} 

\def\keywords#1{{
	\centering{\begin{minipage}{4.5in}\baselineskip=10pt\footnotesize
	{\footnotesize\it Keywords}\/: #1
	\end{minipage}}\par}}

\renewenvironment{thebibliography}[1]			
	{\frenchspacing
	 \ninerm\baselineskip=11pt
	 \begin{list}{\arabic{enumi}.}
	{\usecounter{enumi}\setlength{\parsep}{0pt}
	 \setlength{\leftmargin 12.7pt}{\rightmargin 0pt} 
	 \setlength{\itemsep}{0pt} \settowidth
	{\labelwidth}{#1.}\sloppy}}{\end{list}}

\newcounter{itemlistc}
\newcounter{romanlistc}
\newcounter{alphlistc}
\newcounter{arabiclistc}

\newcommand{\fcaption}[1]{
        \refstepcounter{figure}
        \setbox\@tempboxa = \hbox{\footnotesize Fig.~\thefigure. #1}
        \ifdim \wd\@tempboxa > 5in
           {\begin{center}
        \parbox{5in}{\footnotesize\smalllineskip Fig.~\thefigure. #1}
            \end{center}}
        \else
             {\begin{center}
             {\footnotesize Fig.~\thefigure. #1}
              \end{center}}
        \fi}

\newcommand{\tcaption}[1]{
        \refstepcounter{table}
        \setbox\@tempboxa = \hbox{\footnotesize Table~\thetable. #1}
        \ifdim \wd\@tempboxa > 5in
           {\begin{center}
        \parbox{5in}{\footnotesize\smalllineskip Table~\thetable. #1}
            \end{center}}
        \else
             {\begin{center}
             {\footnotesize Table~\thetable. #1}
              \end{center}}
        \fi}

\def\@citex[#1]#2{\if@filesw\immediate\write\@auxout
	{\string\citation{#2}}\fi
\def\@citea{}\@cite{\@for\@citeb:=#2\do
	{\@citea\def\@citea{,}\@ifundefined
	{b@\@citeb}{{\bf ?}\@warning
	{Citation `\@citeb' on page \thepage \space undefined}}
	{\csname b@\@citeb\endcsname}}}{#1}}

\newif\if@cghi
\def\cite{\@cghitrue\@ifnextchar [{\@tempswatrue
	\@citex}{\@tempswafalse\@citex[]}}
\def\citelow{\@cghifalse\@ifnextchar [{\@tempswatrue
	\@citex}{\@tempswafalse\@citex[]}}
\def\@cite#1#2{{$\null^{#1}$\if@tempswa\typeout
	{IJCGA warning: optional citation argument 
	ignored: `#2'} \fi}}

\def\pmb#1{\setbox0=\hbox{#1}
	\kern-.025em\copy0\kern-\wd0
	\kern.05em\copy0\kern-\wd0
	\kern-.025em\raise.0433em\box0}

\def\fnt#1#2{\footnotetext{\kern-.3em
	{$^{\mbox{\scriptsize #1}}$}{#2}}}

\def\fpage#1{\begingroup
\voffset=.3in
\thispagestyle{empty}\begin{table}[b]\centerline{\footnotesize #1}
	\end{table}\endgroup}

\def\runninghead#1#2{\pagestyle{myheadings}
\markboth{{\protect\footnotesize\it{\quad #1}}\hfill}
{\hfill{\protect\footnotesize\it{#2\quad}}}}
\font\tenrm=cmr10
\font\tenit=cmti10 
\font\tenbf=cmbx10
\font\bfit=cmbxti10 at 10pt
\font\ninerm=cmr9
\font\nineit=cmti9
\font\ninebf=cmbx9
\font\eightrm=cmr8

\def\qed{\hbox{${\vcenter{\vbox{			
   \hrule height 0.4pt\hbox{\vrule width 0.4pt height 6pt
   \kern5pt\vrule width 0.4pt}\hrule height 0.4pt}}}$}}

\def\bsc{{\sc a\kern-6.4pt\sc a\kern-6.4pt\sc a}}	
\def\bflatex{\bf L\kern-.30em\raise.3ex\hbox{\bsc}\kern-.14em 
T\kern-.1667em\lower.7ex\hbox{E}\kern-.125em X} 

\begin{document}

\runninghead{
}{
}

\normalsize\textlineskip
\thispagestyle{empty}
\setcounter{page}{1}



\fpage{1}
\centerline{\bf A SMALL ANGLE NEUTRON SCATTERING STUDY }
\vspace*{0.035truein}
\centerline{\bf OF THE VORTEX MATTER IN La$_{2-x}$Sr$_{x}$CuO$_{4}$ (x=0.17)}
\vspace*{0.37truein}
\centerline{\footnotesize R. GILARDI\footnote{Corresponding author: raffaele.gilardi@psi.ch}, S. STREULE and J. MESOT}
\vspace*{0.015truein}
\centerline{\footnotesize\it Laboratory for Neutron Scattering, ETH Zurich and PSI Villigen} 
\baselineskip=10pt
\centerline{\footnotesize\it CH-5232 Villigen, Switzerland} 
\vspace*{0.1truein}
\centerline{\footnotesize A.J. DREW, U. DIVAKAR and S.L. LEE}
\vspace*{0.015truein}
\centerline{\footnotesize\it School of Physics and Astronomy, University of St. Andrews}
\baselineskip=10pt
\centerline{\footnotesize\it North Haugh, St Andrews KY16 9SS, United Kingdom}
\vspace*{0.1truein}
\centerline{\footnotesize S.P. BROWN and E.M. FORGAN}
\vspace*{0.015truein}
\centerline{\footnotesize\it School of Physics and Astronomy, University of Birmingham}
\baselineskip=10pt
\centerline{\footnotesize\it Birmingham B15 2TT, United Kingdom}
\vspace*{0.1truein}
\centerline{\footnotesize N. MOMONO and M. ODA}
\vspace*{0.015truein}
\centerline{\footnotesize\it Department of Physics, Hokkaido University}
\baselineskip=10pt
\centerline{\footnotesize\it Sapporo 060-0810, Japan}
\vspace*{0.225truein}

\vspace*{0.21truein}
\abstracts{The magnetic phase diagram of slightly overdoped La$_{2-x}$Sr$_{x}$CuO$_{4}$ (x=0.17) is characterised by a field-induced hexagonal to square transition of the vortex lattice at low fields ($\sim$0.4 Tesla) [R. Gilardi \textit{et al., Phys. Rev. Lett.} \textbf{88}, 217003 (2002)]. Here we report on a small angle neutron scattering study of the vortex lattice at higher fields, that reveals no further change of the coordination of the square vortex lattice up to 10.5 Tesla applied perpendicular to the CuO$_2$ planes. Moreover, it is found that the diffraction signal disappears at temperatures well below $T_c$, due to the melting of the vortex lattice.
}{}{}

\vspace*{15pt}
\keywords{small angle neutron scattering; vortex lattice; La$_{2-x}$Sr$_{x}$CuO$_{4}$.}


\vspace*{1pt}\textlineskip	
\vspace*{15pt}

The magnetic phase diagram of the high-temperature superconductors is dominated by the mixed phase, where the magnetic flux can enter the sample in the form of quantised magnetic vortices that interact to form a vortex lattice (VL). Due to a combination of thermal fluctuations and out-of-plane anisotropy effects, the VL can melt into a liquid state above the so-called melting line $B_m(T)$. On the other hand a sufficiently large degree of in-plane anisotropy can affect the symmetry of the VL at low temperatures.
\begin{figure}[htbp]
\begin{center}
\includegraphics[width=12cm]{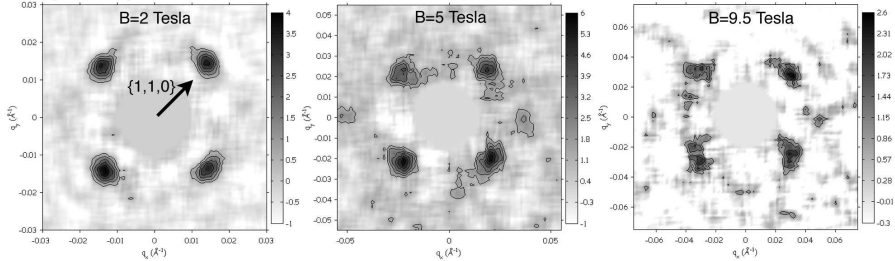}
\end{center}
\caption{SANS diffraction patterns from the vortex lattice in slightly overdoped LSCO, measured at T=5 K at an applied magnetic field of $B$=2 T, 5 T and 9.5 T.}
\end{figure} 

In a previous small angle neutron scattering (SANS) experiment on slightly overdoped La$_{2-x}$Sr$_{x}$CuO$_{4}$ (LSCO) (x=0.17, $T_c$=37 K) we observed a field-induced transition from a hexagonal to a square VL around $B^{*}$=0.4 Tesla\cite{Gilardi}, which is indicative of the coupling of the VL to a source of anisotropy, such as that provided by the d-wave superconducting gap\cite{Shiraishi,Ichioka,Eskildsen} or Fermi surface anisotropies\cite{Nakai}.
In LSCO the position of the gap nodes and of the Fermi velocity minima are rotated 45$^o$ with respect to each other. There thus exists the potential for a competition between a square VL oriented in reciprocal space along the node directions (\{1,0,0\} directions in orthorhombic LSCO) and one oriented along the Cu-O bonds (\{1,1,0\} directions). For  $B^{*}<B<1.2$ T, the square VL is oriented  along the \{1,1,0\} directions\cite{Gilardi}, but a reorientation (rotation by 45$^o$) of the square VL has been predicted to occur at higher fields\cite{Nakai,Machida}. 
In order to test this possibility, we have extended our measurements to higher magnetic fields. Moreover, we have also investigated the temperature dependence of the VL scattered intensity. 

The experiments were carried out using the SANS-I diffractometer at the Paul Scherrer Institute, Switzerland.
The LSCO single crystal was mounted in a horizontal field cryomagnet with the c-axis along the neutron beam. The magnetic field (up to 10.5 Tesla) was aligned parallel to the neutron beam and therefore perpendicular to the CuO$_2$ planes. In Fig.1 we show SANS diffraction patterns obtained at T=5 K after field-cooling in $B$=2 T, 5 T and 9.5 T.  A background measured at $T$=40 K (above $T_c$) has been subtracted in order to remove the large signal arising from crystal defect scattering. The diffracted intensity is  concentrated in four bright spots forming a perfect square. The square VL is always oriented along the \{1,1,0\} directions (Cu-O bonds), therefore indicating no reorientation of the VL up to the highest magnetic field applied (10.5 Tesla).

The temperature dependence of the scattered neutron intensity at $B$=5 T, summed over all diffraction peaks, is shown in Fig.2a. As a function of temperature, the intensity rapidly decreases and finally vanishes at $T_m$ well below $T_{c2}$, where the normal state is recovered. 
$T_m$ is usually interpreted as the melting temperature of the VL. Interestingly, the scattered intensity is negligible above $T_m$, similarly to what has been observed in Bi$_{2.15}$Sr$_{1.95}$CaCu$_2$O$_{8+x}$\cite{Cubitt} and YBa$_2$Cu$_3$O$_{7-x}$\cite{Aegerter}. 
\begin{figure}[htbp]
\begin{center}
\includegraphics[width=12cm]{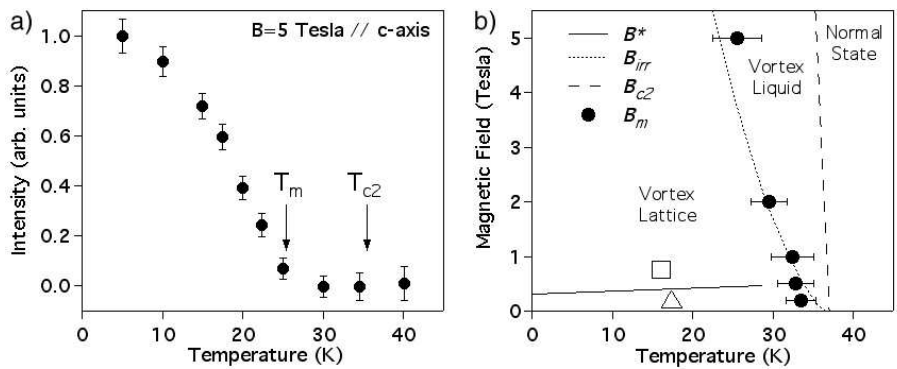}
\end{center}
\caption{a) Temperature dependence of the diffracted intensity measured at $B$=5 T. b) Magnetic phase diagram of LSCO (x=0.17). The melting line $B_m$ (filled circles) has been determined by SANS measurements. $B_{irr}$ (dotted line) and $B_{c2}$ (dashed line) have been obtained by macroscopic measurements, see text. 
The hexagonal ($\triangle$) to square ($\Box$) transition line $B^{*}$ is schematically shown by the solid line.}
\end{figure}
This is surprising, since in a liquid of straight vortices a ring of intensity is expected.
This may indicate that the vortices are entangled (or decoupled)  in the liquid phase.
The values obtained for $B_m(T)$ are plotted in the magnetic phase diagram shown in Fig.2b together with the results of macroscopic measurements\cite{Gilardi2}. The upper critical field $B_{c2}(T)$, indicating  the crossover line to the normal state, has been determined using the Landau-Ott scaling procedure\cite{Landau} for magnetization data. The irreversibility line $B_{irr}(T)$ has been obtained from the loss peak in the imaginary part of the ac-susceptibility and is associated with the onset of resistivity and reversible magnetization. The nice agreement between the melting line $B_m(T)$ as determined by SANS, and the irreversibility line $B_{irr}(T)$ should be noted.

\nonumsection{Acknowledgements}
\noindent
This work was performed at the spallation neutron source SINQ, Paul Scherrer Institut, Villigen, Switzerland. The expert technical assistance by J. Kohlbrecher is gratefully acknowledged. This work was supported by the Swiss National Science Foundation, the Engineering  and Physical Science Research Council of  the U.K., and the Ministry of Education  and Science of Japan.

\nonumsection{References}


\begin{thebibliography}{000}
\bibitem{Gilardi2}
These measurements have been performed on a small piece of the single crystal used in the SANS experiments and will be published elsewhere.
\end{thebibliography}
\end{document}